\documentclass[10pt,conference]{IEEEtran}
\IEEEoverridecommandlockouts
\usepackage{cite}
\usepackage{verbatim} 
\usepackage{amsmath,amssymb,amsfonts}
\usepackage{algorithmic}
\usepackage[linesnumbered,ruled]{algorithm2e}
\usepackage{tabularx}
\usepackage{graphicx}
\usepackage{textcomp}
\usepackage{xcolor}
\usepackage{accents}
\SetKwInput{KwInput}{Input}                
\SetKwInput{KwOutput}{Output}              
\def\BibTeX{{\rm B\kern-.05em{\sc i\kern-.025em b}\kern-.08em
    T\kern-.1667em\lower.7ex\hbox{E}\kern-.125emX}}

\begin{document}

\newcommand{\force}{\mbox{$\Vdash$}}

\newcommand{\ghat}{\mbox{$\bm \hat g$}}
\newcommand{\Nhat}{\mbox{$\bm {\hat N}$}}
\newcommand{\Nlow}{\mbox{${N_{\hbox{\tiny low}}}$}}
\newcommand{\Nhigh}{\mbox{${N_{\hbox{\tiny high}}}$}}
\newcommand{\Nmax}{\mbox{${N_{\hbox{\tiny max}}}$}}
\newcommand{\Iul}{\mbox{${I_{\hbox{\tiny ul}}}$}}

\newcommand{\va}{\mbox{${\bf a}$}}
\newcommand{\vah}{\mbox{${\bf \hat a}$}}
\newcommand{\vat}{\mbox{${\bf \tilde a}$}}
\newcommand{\vb}{\mbox{${\bf b}$}}
\newcommand{\vc}{\mbox{${\bf c}$}}

\newcommand{\vp}{\mbox{${\bf p}$}}
\newcommand{\vd}{\mbox{${\bf d}$}}
\newcommand{\rh}{\mbox{${\hat r}$}}

\newcommand{\vx}{\mbox{${\bf x}$}}
\newcommand{\vxt}{\mbox{${\bf \tilde x}$}}
\newcommand{\vh}{\mbox{${\bf h}$}}
\newcommand{\htld}{\mbox{${\tilde h}$}}
\newcommand{\vhh}{\mbox{${\bf \hat h}$}}
\newcommand{\hhat}{\mbox{${\hat h}$}}
\newcommand{\ve}{\mbox{${\bf e}$}}
\newcommand{\vg}{\mbox{${\bf g}$}}
\newcommand{\vgh}{\mbox{${\bf \hat g}$}}
\newcommand{\vq}{\mbox{${\bf q}$}}
\newcommand{\vt}{\mbox{${\bf t}$}}
\newcommand{\vw}{\mbox{${\bf w}$}}
\newcommand{\vdw}{\mbox{${\bf dw}$}}

\newcommand{\vwh}{\mbox{${\bf \hat w}$}}
\newcommand{\wh}{\mbox{${\hat w}$}}
\newcommand{\vwt}{\mbox{${\bf \tilde w}$}}
\newcommand{\wt}{\mbox{${\tilde w}$}}
\newcommand{\vs}{\mbox{${\bf s}$}}
\newcommand{\vsh}{\mbox{${\bf \hat s}$}}
\newcommand{\vst}{\mbox{${\bf \tilde s}$}}
\newcommand{\vr}{\mbox{${\bf r}$}}
\newcommand{\vv}{\mbox{${\bf v}$}}
\newcommand{\vu}{\mbox{${\bf u}$}}
\newcommand{\vy}{\mbox{${\bf y}$}}
\newcommand{\vz}{\mbox{${\bf z}$}}
\newcommand{\vn}{\mbox{${\bf n}$}}
\newcommand{\vnt}{\mbox{${\bf \tilde n}$}}
\newcommand{\ntld}{\mbox{${\tilde n}$}}
\newcommand{\nhat}{\mbox{${\hat n}$}}
\newcommand{\phat}{\mbox{${\hat p}$}}
\newcommand{\vzero}{\mbox{${\bf 0}$}}
\newcommand{\vone}{\mbox{${\bf 1}$}}

\newcommand{\mA}{\hbox{{\bf A}}}
\newcommand{\mAh}{\mbox{${\bf \hat A}$}}
\newcommand{\mAt}{\mbox{$\bf \tilde A$}}
\newcommand{\mB}{\hbox{{\bf B}}}
\newcommand{\mBh}{\mbox{${\bf \hat B}$}}
\newcommand{\mC}{\hbox{{\bf C}}}
\newcommand{\mCh}{\mbox{${\bf \hat C}$}}
\newcommand{\mD}{\hbox{{\bf D}}}
\newcommand{\mDt}{\mbox{$\bf \tilde D$}}
\newcommand{\mE}{\hbox{{\bf E}}}
\newcommand{\mG}{\hbox{{\bf G}}}
\newcommand{\mF}{\hbox{{\bf F}}}
\newcommand{\mH}{\hbox{{\bf H}}}
\newcommand{\mHb}{\mbox{${\bf \bar H}$}}
\newcommand{\mHt}{\mbox{${\bf \tilde H}$}}
\newcommand{\mI}{\hbox{{\bf I}}}
\newcommand{\mJ}{\mbox{${\mathbf J}$}}

\newcommand{\mIh}{\mbox{${\bf \hat I}$}}
\newcommand{\mN}{\hbox{{\bf N}}}
\newcommand{\mM}{\hbox{{\bf M}}}
\newcommand{\mMh}{\mbox{{$\bf \hat M$}}}
\newcommand{\mP}{\hbox{{\bf P}}}
\newcommand{\mQ}{\hbox{{\bf Q}}}
\newcommand{\mR}{\mbox{{$\bf R$}}}
\newcommand{\mRh}{\mbox{{$\bf \hat R$}}}
\newcommand{\mRt}{\mbox{{$\bf \tilde R$}}}
\newcommand{\mS}{\mbox{{$\bf S$}}}
\newcommand{\mSb}{\mbox{{$\bf \bar S$}}}
\newcommand{\mSh}{\mbox{{$\bf \hat S$}}}
\newcommand{\mSt}{\mbox{{$\bf \tilde S$}}}
\newcommand{\mT}{\mbox{{$\bf T$}}}
\newcommand{\mU}{\mbox{{$\bf U$}}}
\newcommand{\mUh}{\mbox{{$\bf \hat U$}}}
\newcommand{\mV}{\mbox{{$\bf V$}}}
\newcommand{\mVh}{\mbox{{$\bf \hat V$}}}
\newcommand{\mW}{\hbox{{\bf W}}}
\newcommand{\mWa}{\mbox{${\bf W_{\ga}}$}}

\newcommand{\mWh}{\mbox{${\bf \hat W}$}}
\newcommand{\mWt}{\mbox{${\bf \tilde W}$}}
\newcommand{\mX}{\mbox{{$\bf X$}}}
\newcommand{\mY}{\mbox{{$\bf Y$}}}
\newcommand{\mZ}{\mbox{{$\bf Z$}}}


\newcommand{\ga}{\alpha}
\newcommand{\gb}{\beta}
\newcommand{\grg}{\gamma}
\newcommand{\gd}{\delta}
\newcommand{\gre}{\varepsilon}
\newcommand{\gep}{\epsilon}
\newcommand{\gz}{\zeta}
\newcommand{\gzh}{\mbox{$ \hat \zeta$}}
\newcommand{\gh}{\eta}
\newcommand{\gth}{\theta}
\newcommand{\gthh}{\mbox{$ \hat \theta$}}
\newcommand{\gi}{iota}
\newcommand{\gk}{\kappa}
\newcommand{\gl}{\lambda}
\newcommand{\gm}{\mu}
\newcommand{\gn}{\nu}
\newcommand{\gx}{\xi}
\newcommand{\gp}{\pi}
\newcommand{\gph}{\phi}
\newcommand{\gr}{\rho}
\newcommand{\gs}{\sigma}
\newcommand{\gsh}{\hat \sigma}
\newcommand{\gt}{\tau}
\newcommand{\gu}{\upsilon}
\newcommand{\gf}{\varphi}
\newcommand{\gc}{\chi}
\newcommand{\go}{\omega}
\newcommand{\ebj}{e^{j \omega}}
\newcommand{\ebjz}{e^{-j \omega}}

\newcommand{\gG}{\Gamma}
\newcommand{\gD}{\Delta}
\newcommand{\gTh}{\Theta}
\newcommand{\gL}{\Lambda}
\newcommand{\gX}{\Xi}
\newcommand{\gP}{\Pi}
\newcommand{\gS}{\Sigma}
\newcommand{\gU}{\Upsilon}
\newcommand{\gF}{\Phi}
\newcommand{\gO}{\Omega}


\def\bm#1{\mbox{\boldmath $#1$}}
\newcommand{\vga}{\mbox{$\bm \alpha$}}
\newcommand{\vgb}{\mbox{$\bm \beta$}}
\newcommand{\vgd}{\mbox{$\bm \delta$}}
\newcommand{\vge}{\mbox{$\bm \epsilon$}}
\newcommand{\vgl}{\mbox{$\bm \lambda$}}
\newcommand{\vgr}{\mbox{$\bm \rho$}}
\newcommand{\vgrh}{\mbox{$\bm \hat \rho$}}
\newcommand{\vgrt}{\mbox{$\bm {\tilde \rho}$}}
\newcommand{\vgp}{\mbox{$\bm \pi$}}

\newcommand{\vgt}{\mbox{$\bm \gt$}}
\newcommand{\vgn}{\mbox{$\bm \gn$}}
\newcommand{\vgth}{\mbox{$\bm {\hat \tau}$}}
\newcommand{\vgtt}{\mbox{$\bm {\tilde \tau}$}}
\newcommand{\vpsi}{\mbox{$\bm \psi$}}
\newcommand{\vphi}{\mbox{$\bm \phi$}}
\newcommand{\vxi}{\mbox{$\bm \xi$}}
\newcommand{\vth}{\mbox{$\bm \theta$}}
\newcommand{\vthh}{\mbox{$\bm {\hat \theta}$}}

\newcommand{\mgG}{\mbox{$\bm \Gamma$}}
\newcommand{\mgGh}{\mbox{$\hat {\bm \Gamma}$}}
\newcommand{\mgD}{\mbox{$\bm \Delta$}}
\newcommand{\mgDone}{\mbox{$\bm \Delta^{(1)}$}}
\newcommand{\mgDtwo}{\mbox{$\bm \Delta^{(2)}$}}

\newcommand{\mgU}{\mbox{$\bm \Upsilon$}}
\newcommand{\mgL}{\mbox{$\bm \Lambda$}}
\newcommand{\mPsi}{\mbox{$\bm \Psi$}}
\newcommand{\mgX}{\mbox{$\bm \Xi$}}
\newcommand{\mgS}{\mbox{$\bm \Sigma$}}

\newcommand{\oA}{{\open A}}
\newcommand{\oC}{{\open C}}
\newcommand{\oF}{{\open F}}
\newcommand{\oN}{{\open N}}
\newcommand{\oP}{{\open P}}
\newcommand{\oQ}{{\open Q}}
\newcommand{\oR}{{\open R}}
\newcommand{\oZ}{{\open Z}}


\newcommand{\Nu}{{\cal V}}
\newcommand{\cA}{{\cal A}}
\newcommand{\cB}{{\cal B}}
\newcommand{\cC}{{\cal C}}
\newcommand{\cD}{{\cal D}}
\newcommand{\cF}{{\cal F}}
\newcommand{\cH}{{\cal H}}
\newcommand{\cK}{{\cal K}}
\newcommand{\cI}{{\cal I}}
\newcommand{\cL}{{\cal L}}
\newcommand{\cM}{{\cal M}}
\newcommand{\cN}{{\cal N}}
\newcommand{\cO}{{\cal O}}
\newcommand{\cP}{{\cal P}}
\newcommand{\cR}{{\cal R}}
\newcommand{\cS}{{\cal S}}
\newcommand{\cU}{{\cal U}}
\newcommand{\cV}{{\cal V}}
\newcommand{\cT}{{\cal T}}
\newcommand{\cX}{{\cal X}}

\newcommand{\rH}{^{*}}
\newcommand{\rT}{^{ \raisebox{1.2pt}{$\rm \scriptstyle T$}}}
\newcommand{\rF}{_{ \raisebox{-1pt}{$\rm \scriptstyle F$}}}
\newcommand{\rE}{{\rm E}}

\newcommand{\dom}{\hbox{dom}}
\newcommand{\rng}{\hbox{rng}}
\newcommand{\Span}{\hbox{span}}
\newcommand{\Ker}{\hbox{Ker}}
\newcommand{\On}{\hbox{On}}
\newcommand{\otp}{\hbox{otp}}
\newcommand{\ZFC}{\hbox{ZFC}}
\def\Re{\ensuremath{\hbox{Re}}}
\def\Im{\ensuremath{\hbox{Im}}}
\newcommand{\SNR}{\ensuremath{\hbox{SNR}}}
\newcommand{\SINR}{\ensuremath{\hbox{SINR}}}
\newcommand{\CRB}{\ensuremath{\hbox{CRB}}}
\newcommand{\diag}{\ensuremath{\hbox{diag}}}
\newcommand{\trace}{\ensuremath{\hbox{tr}}}

\newcommand{\dlot}{\mbox{$\delta^1_3$}}
\newcommand{\Dlot}{\mbox{$\Delta^1_3$}}
\newcommand{\Dlof}{\mbox{$\Delta^1_4$}}
\newcommand{\dlof}{\mbox{$\delta^1_4$}}
\newcommand{\bP}{\mbox{$\bf{P}$}}
\newcommand{\Pot}{\mbox{$\Pi^1_2$}}
\newcommand{\Sot}{\mbox{$\Sigma^1_2$}}
\newcommand{\gDot}{\mbox{$\gD^1_2$}}

\newcommand{\Potr}{\mbox{$\Pi^1_3$}}
\newcommand{\Sotr}{\mbox{$\Sigma^1_3$}}
\newcommand{\gDotr}{\mbox{$\gD^1_3$}}

\newcommand{\Pofr}{\mbox{$\Pi^1_4$}}
\newcommand{\Sofr}{\mbox{$\Sigma^1_4$}}
\newcommand{\Dofr}{\mbox{$\gD^1_4$}}

\newcommand{\Sa}{\mbox{$S_{\ga}$}}
\newcommand{\Qk}{\mbox{$Q_{\gk}$}}
\newcommand{\Ca}{\mbox{$C_{\ga}$}}

\newcommand{\gkp}{\mbox{$\gk^+$}}
\newcommand{\aron}{ Aronszajn }

\newcommand{\sqkp}{\mbox{$\Box_{\gk}$}}
\newcommand{\dkp}{\mbox{$\Diamond_{\gk^{+}}$}}
\newcommand{\sqsqnce}
{\mbox{\\ $\ < \Ca \mid \ga < \gkp \ \ \wedge \ \ \lim \ga >$ \ \ }}
\newcommand{\dsqnce}{\mbox{$<S_{\ga} \mid \ga < \gkp >$}}

\newtheorem{theorem}{Theorem}
\newtheorem{lemma}[theorem]{Lemma}
\newtheorem{prop}{Proposition}
\newtheorem{claim}[theorem]{Claim}

\newtheorem{definition}{Definition}
\newtheorem{question}{Question}
\newtheorem{coro}{Corollary}

\newcommand{\beq}{\begin{equation}}
\newcommand{\eeq}{\end{equation}}
\newcommand{\bea}{\begin{array}}
\newcommand{\ena}{\end{array}}
\newcommand{\bds}{\begin {description}}
\newcommand{\eds}{\end {description}}
\newcommand{\bdf}{\begin{definition}}
\newcommand{\blm}{\begin{lemma}}
\newcommand{\edf}{\end{definition}}
\newcommand{\elm}{\end{lemma}}
\newcommand{\bthm}{\begin{theorem}}
\newcommand{\ethm}{\end{theorem}}
\newcommand{\bprp}{\begin{prop}}
\newcommand{\eprp}{\end{prop}}
\newcommand{\bcl}{\begin{claim}}
\newcommand{\ecl}{\end{claim}}
\newcommand{\bcr}{\begin{coro}}
\newcommand{\ecr}{\end{coro}}
\newcommand{\bquest}{\begin{question}}
\newcommand{\equest}{\end{question}}

\newcommand{\rarrow}{{\rightarrow}}
\newcommand{\Rarrow}{{\Rightarrow}}
\newcommand{\larrow}{{\leftarrow}}
\newcommand{\Larrow}{{\Leftarrow}}
\newcommand{\restrict}{{\upharpoonright}}
\newcommand{\nin}{{\not \in}}



\newcommand{\ie}{\hbox{i.e.}}
\newcommand{\eg}{\hbox{e.g.}}
\newcommand{\Pnx}{\mbox{$P_{\hbox{next}}$}}
\newcommand{\Knx}{\mbox{$K_{\hbox{next}}$}}

\title{Distributed Deep Reinforcement Learning  for Collaborative Spectrum Sharing}
\author{
  \IEEEauthorblockN{Pranav M. Pawar\thanks{This study was partially supported by ISF grants 2277/16 and 1644/18.
Pranav M. Pawar is also supported by an Israeli Planning and Budget
Committee (PBC) post-doctoral fellowship (2019-2021). Pranav M. Pawar
was with the Faculty of Engineering, Bar-Ilan univer-sity, Ramat-Gan, 52900,
Israel. Currently, he is with the Dept. of Com-puter Science, BITS Pilani,
Dubai-345055, UAE (email: pranav@dubai.bits-pilani.ac.in).Amir Leshem is
with the Faculty of Engineering, Bar-Ilan university,Ramat-Gan, 52900, Israel
(email: leshema@biu.ac.il).}}   \IEEEauthorblockA{Dept.  of  Computer Science \\ BITS Pilani, Dubai-345055, UAE}
  \and
  \IEEEauthorblockN{Amir Leshem}
  \IEEEauthorblockA{Faculty  of  Engineering \\ Bar-Ilan  university,  Ramat-Gan,  52900,  Israel}

}

\maketitle

\begin{abstract}
Spectrum sharing among users is a fundamental
problem in the management of any wireless network. In this paper, we discuss the problem of distributed spectrum collaboration
without central management under general unknown channels.
Since the cost of communication, coordination and control is
rapidly increasing with the number of devices and the expanding
bandwidth used there is an obvious need to develop distributed
techniques for spectrum collaboration where no explicit signaling 
is used. In this paper, we combine game-theoretic insights
with deep Q-learning to provide a novel asymptotically optimal
solution to the spectrum collaboration problem. We propose a
deterministic distributed deep reinforcement learning (D3RL)
mechanism using a deep Q-network (DQN). It chooses the channels using the Q-values and the channel loads while limiting the
options available to the user to a few channels with the highest Qvalues and among those, it selects the least loaded channel. Using
insights from both game theory and combinatorial optimization
we show that this technique is asymptotically optimal for large
overloaded networks. The selected channel and the outcome of
the successful transmission are fed back into the learning of
the deep Q-network to incorporate it into the learning of the Qvalues. We also analyzed performance to understand the behavior
of D3RL in different conditions and compared it to state-of-theart techniques.
\end{abstract}


\section{Introduction}
\label{intro}
Spectrum sharing is a fundamental problem for the efficient management of wireless communication networks. There are two paradigms for network management: centralized and distributed. The former is prevalent in cellular networks, while the latter has been mainly used in ad-hoc networks. However, the growing demand for bandwidth, the increase in the number of users, and the scarcity of the available spectrum make collecting channel state information increasingly hard \cite{Duan2014}. In this paper, we approach the problem of spectrum sharing by combining insights from our previous game-theoretic analysis of the interference game \cite{Bistritz&Leshem2019} with deep $Q$-learning. This enables us to boost the performance of deep $Q$-learning by $20-50 \%$ compared to the algorithm in \cite{Naparstek&Cohen2019} in overloaded settings when the load per frequency channel is larger than $1$. 

The most suitable and efficient way to manage channels in this type of scenario is to learn the channel behavior (state and action profile) and allocate the channel accordingly. Hence, developing distributed learning solutions for the channel allocation problem has attracted major attention in recent years. This paper aims to develop a distributed learning algorithm for channel allocation for real-time multi-user networks, without the exchange of multiple messages or large state information. We adopted a deep reinforcement learning (DRL) \cite{Hasselt2016} mechanism to achieve this goal because it gives a good approximation of objective values. The objective of DRL is to learn efficient strategies and rules for a given decision problem. Here, we used the most popular reinforcement learning algorithm, $Q$-learning, which is combined with a deep neural network; i.e., Deep $Q$-network (DQN) \cite{Hasselt2016}. The DQN is used for mapping the state with actions to maximize a $Q$-value.

The basic drawback of multi-agent $Q$-learning is that even if convergence is achieved, it is only to a Nash Equilibrium (NE) point. As is well-known a NE can be highly sub-optimal in terms of social welfare and fairness \cite{leshem2009}. Moreover, by modifying the utilities used for optimization by the agents and allowing each agent to select only a few best strategies produces a game where all the NE points are indeed near-optimal (in terms of the original utility)\cite{Bistritz&Leshem2019}. This suggests that a similar modification to the Q-learning process might yield better allocations. In this paper, we use the basic idea of balanced allocation together with the classical result reported in \cite{Azar2000} to obtain asymptotic optimal learning rules. 

Our main contribution in this paper is a modification of the  architecture in \cite{Naparstek&Cohen2019} by modifying the allocation rules and improving and changing the game used for the multi-agent Q-learning by exploiting the game theoretic result of \cite{Bistritz&Leshem2019}. First, we limit the strategy space to the  $M$ (actually $M=2$ is sufficient) channels with the highest $Q$-values. This results in a game where all NE points are near optimal from a social welfare perspective.  Next, we use a deterministic strategy to choose the least loaded channel among these $M$ best channels. This yields a near-optimal allocation in the following sense
\begin{itemize}
    \item Each user obtains one of its $M$ best channels. so by
    order statistics, it is asymptotically optimal, as long as $\frac{M}{N} \to 0, N \to \infty$ and the fading statistics is exponentially dominated (an assumption which holds for all standard fading models).
    \item The load on all channels is nearly equal; i.e., $\frac{N}{K} + O(\frac {\log \log K}{\log M})$.

\end{itemize}
The intuition behind this second step, is that we consider all the best $M$ channels as having the same utility which results in an equivalent game with only good equilibria. The proof of these two results will be given the full version of the paper. It combines ideas from \cite{Bistritz&Leshem2019} and \cite{Azar2000}. 
\section{Related Work}
\label{RW}
Many recent developments in distributed learning methods have demonstrated that it can be an efficient technique to solve the spectrum (or channel) sharing problems in wireless networks. This section discusses the state-of-the-art in the area of distributed learning methods for spectrum sharing.

In \cite{Naparstek&Cohen2019} a deep $Q$-learning for spectrum access (DQSA) is discussed. Here, spectrum actions for every user are learned through training a DQN. The algorithm learns good strategies for every user online and in a distributed fashion without exchanging messages or with online coordination among users. In \cite{Nguyen2018} the authors put forward an optimal access policy
using the state transition probabilities by considering Markov channels. It also proposes an optimal access policy for these channels using deep $Q$-learning if the transition probabilities are unknown. Here, the DQN uses $\varepsilon$-greedy policies to accumulate the training data. Spectrum resource allocation in a cognitive radio network is explored in \cite{Li2018}. It concentrates on developing a deep reinforcement learning (DRL) mechanism for power control of the secondary user to share the spectrum with a primary user. The $Q$-learning-based DRL is implemented by the secondary users to learn and adjust transmit power after interactions with the primary users so that both users can transmit successfully. In \cite{Zafaruddin2019} distributed learning for the channel, allocation considers the multi-arm bandit scenario, which achieves an optimal regret of $O(\log T)$. It describes a distributed channel allocation mechanism that assumes carrier sense multiple access (CSMA). Here, the user learns from observing a single channel without decoding a channel and without exchanging extra
information between users. An online self-decision and offline self-learning algorithms are proposed for channel allocation in \cite{QiaoMu2017}, for multi-channel wireless sensor network (WSN). A non-cooperative game is used for online self-decision algorithms and a $Q$-learning-based DRL approach is used for offline self-learning of channels. The findings show that offline self-learning converges to optimal channel selection with lower computational and storage resources. In \cite{Wang2018} a channel access application of DQN is considered in a multi-user, multi-hop, and simultaneous transmission scenario of WSN. It suggests a DQN to access channel using online learning. The DQN approach considers a large system and finds an optimal policy from historical observations without prior knowing system dynamics. Deep-Reinforcement Learning Multiple Access (DLMA) \cite{YYu2019} is a heterogeneous MAC protocol using
DRL. Here, the DRL agent learns an optimal medium access control (MAC) strategy for efficient co-existence with time division multiple access (TDMA) and ALOHA nodes. This DRL learns through a series of state-action-rewards. This work also concentrates on analyzing the characteristics of DRL as compared to other neural network techniques to then apply it to wireless networks. The sensor scheduling problem for wireless channel allocation using DRL was studied in \cite{Leong2020} where the  scheduling problem, it is formulated as a Markov Decision Process (MDP) and solved using DQN. The authors report good performance as compared to other sub-optimal sensor scheduling policies. Finally, these studies are primarily focused on non-deterministic solutions for spectrum access
using Q-learning which is more complex in implementation and finding optimal values in real-world conditions.

The remainder of this paper is organized as follows. Section III presents the network model for the proposed technique and formulates the problem. Section IV describes the D3RL mechanism with its architecture, algorithm, and working. Section V presents the communication- and neural network parameters for simulation and analyzes the simulation results. Section VI concludes the paper and outlines future work.

\section{Network Model and Problem Statement}
\label{NMandPS}
We consider an ad-hoc network with $N$ users and $K$ channels, where $N>K$. The training is performed in a distributed manner at each user. For simplicity of exposition, we assume that all users know both $N$ and $K$ (This can be broadcasted to all users if needed). Let $M$ be a parameter designating a small set of best channels for transmission, $ M \ll K$. All users in the network have similar capabilities. All nodes in the network are synchronized as is typical in slotted random access networks.

We assume a slotted random access mechanism for sharing the spectrum; specifically, we consider slotted-multi-channel ALOHA transmission where each user is allowed to transmit on a specific channel in a specific slot according to a certain transmission probability \cite{Cohen2016}. Here, each user transmits with probability $p_T=\frac{K}{N}$ and does not transmit at a probability of $1-p_T$.  We do not assume that users know a-priori the channel qualities or the loads on the channel and our goal is to devise a multi-agent learning algorithm that will lead the network to an allocation which is good for all users where each user transmits over one of its $M$ best channels and the load on all channels is approximately identical. 

To analyze the network, we assume that at all times each user has pure action set $a_n \in \left\{0,\ldots,K\right\}$ and transmits over channel $a_n$ using a slotted ALOHA protocol. For loaded ALOHA we will mostly depend on a mixed strategy of the players. A mixed strategy for a player is a probability distribution over the possible channels (pure strategy) $\gs_n(t)=\left[p_{n,0}(t),\ldots,p_{n,K}(t)\right]$, where $p_{n,0}(t)$ is the probability of not transmitting at time $t$, and $p_{n,k}(t)$ is the probability of user $n$  transmitting over channel $k$ at time $t$ for $k>0$. 

Let $o_n(t)$ be a binary observation indicating whether a packet is successfully delivered or not; \ie,  $1$ if acknowledgment (ACK) is received and $0$ otherwise. Let $a_n(t)$ be the action of user $n$  at time $t$ and $\sigma_n(t)$ be the strategy of user $n$ at time $t$.

Let $r_n(t)$ be the reward that $n$ obtains at time $t$, $u_n$ be the utility of user and $L_k$ is the load on channel $k$.  

\textit{Definition 1:} A history of user $n$ at time $t$ is the set of all actions, observations, and load on channels up to time $t$ is defined as
\begin{equation}
\label{def:History}
H_{n}(t) = \bigg(\{{a_n}(i)\}_{i=1}^t,\{o_n(i)\}_{i=1}^t, \{L_i\}_{i=1}^t\bigg)
\end{equation}
History $H_{n}(t)$ is used for training the users to learn their best strategy. 

\textit{Definition 2:} The utility function (an instantaneous functions) that defines the throughput of user $n$ on channel $k$ is
\begin{equation}
\label{def:Util}
u_n(\underaccent{\bar}{a})= \begin{cases}
B\log_2(1+\frac{P_n\mid (h_n(k)\mid^2}{\sigma_n^2}),& \substack{a_n = k, \\ a_m\neq k, 
\forall m \neq n} \\
0,&{otherwise}
\end{cases}
\end{equation}
Here, $P_n$ is the users transmission power, $h_n$ is the channel gain, $B$ is the bandwidth of each channel and $\underaccent{\bar}{a}$ is a vector of actions (where $a_n$ is the action of user $n$, i.e., the frequency it selected.  

\textit{Definition 3:} The  multi-channel random access game is defined by:
\begin{equation}
\label{def:Game}
G= \bigg<N,\{0\ldots,K\}^n,\left(u_n:n=1\ldots,N\right)\bigg>,
\end{equation}
where $A=\left\{0,1,...,K \right\}$ is the set of actions of each player. $0$ denotes no transmission, while $0<k\leq K$ denotes the identity of the selected channel for transmission.

\textit{Definition 4:} A mixed strategy is a probability distribution with respect to pure strategy $\sigma$, which is 
\begin{equation}
    \label{def:MSt}
    \Bar{u_n}(\underaccent{\bar}{\sigma}) = \mathbb E_{\underaccent{\bar}{\sigma}}(u_n(a_n)) = \sum_{\underaccent{\bar}{a} \epsilon \{0,\ldots,K\}^n}\Big(\prod_{n=1}^N\sigma_n(a_n)\Big)u_n(\underaccent{\bar}{a})
\end{equation}
Here, $\underaccent{\bar}{\sigma}$ is a strategy vector and $\Bar{u_n}(\underaccent{\bar}{\sigma})$ is a payoff for the considered strategy vector.

\textit{Definition 5:} The total accumulated reward with discount factor $\grg$ for player $n$, $R_n$ is given by: 
\begin{equation}
\label{def:AccR}
R_n = \sum_{t=1}^T \gamma^{t-1}  u_n(\underaccent{\bar}{\sigma})
\end{equation}
$0 \leq \gamma \leq 1$ and $T$ is the time-horizon.

\textit{Definition 6:} The strategy profile $(\sigma_n^*, \sigma_{-n}^*)$ is called a NE in the multi-channel random access game if 
\begin{equation}
\label{def:NE}
    R_{n}(\sigma_{n}^*, \sigma_{-n}^*) \geq R_{n}(\sigma_n,  \sigma_{-n}^* )
\end{equation}
for all $\sigma_n$ and all $n \in N$. Here, $\sigma_{-n}$ is a strategy profile for all users except user $n$.

The objective is to find a strategy $\sigma_n$ for a user $n$, which maximizes the expected accumulated discounted reward, $\max_{\sigma_n}  \mathbb E [R_n(\sigma_n,\sigma_{-n})]$.

A NE is a stable point in the dynamics, hence it is desired to achieve such equilibrium which will prevent network fluctuations. However, in general, such equilibria can be highly sub-optimal \cite{leshem2009}. However, recently Bistritz and Leshem \cite{Bistritz&Leshem2019} proved that by changing the utility or restricting the strategies utilized by each player a competitive game can be formed where each NE point is near-optimal with respect to the sum rate of all users.

In this paper, we develop a learning strategy that always achieves a NE with high utility for all users. To that end, we exploit the results of \cite{Bistritz&Leshem2019} to modify the DQN learning \cite{Mnih2015}. This is successfully achieved using DRL techniques, $Q$-learning \cite{Watkins92q-learning}, DQN \cite{Mnih2015} and double $Q$-learning \cite{Hasselt2016}.

\section{Deterministic Distributed Deep Reinforcement Learning}
\label{D3RL}
In this section, we present a new collaborative spectrum access technique called Deterministic, Distributed Deep Reinforcement Learning (D3RL). The basic idea is to deterministically limit the set of strategies of each player and to enforce users to use the least loaded channels among their best $M$ strategies. The technique implements a deep reinforcement network for learning the strategies and a decision mechanism on the transmitted channel which is deterministic, given the computed $Q$-values. The first subsection describes the architecture, then the operation phase and load estimation in the algorithm. Finally, we discuss the training of the algorithm.
\subsection{Architecture}
\begin{figure}
\center{\includegraphics[width=0.45\textwidth]{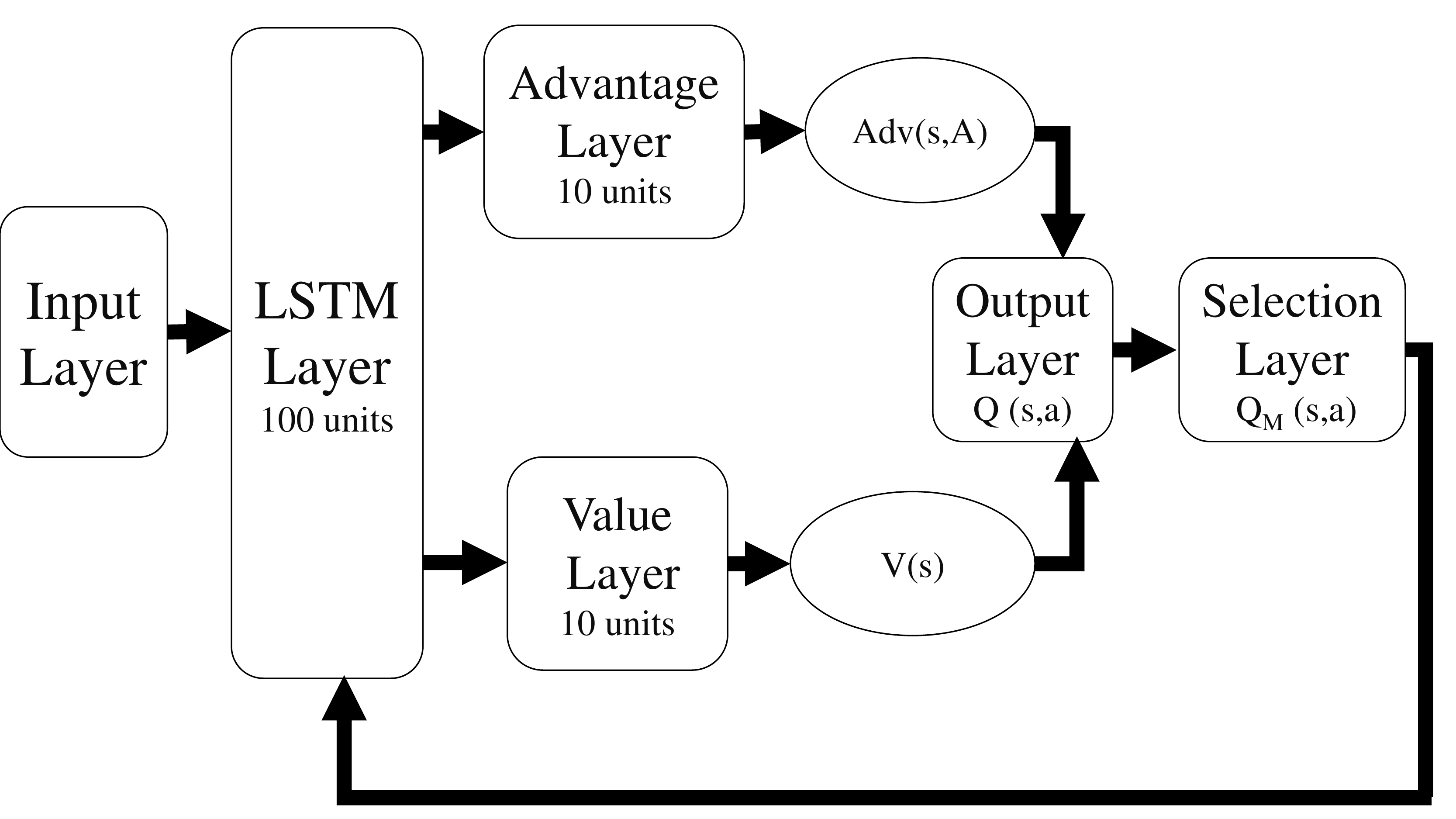}}
\caption{Architecture of the DQN used in D3RL}
\label{fig:Arch}
\end{figure}
This subsection describes the proposed layered architecture for the reinforcement learning used in D3RL to solve the channel allocation problem in the multi-user network. The architecture of D3RL consists of an input layer, long short term memory (LSTM) layers, value layers, advantage layers, an output layer, a selection layer, and distributed double $Q$-learning. Here, D3RL chooses the channel in a deterministic way by using $Q$-values and the load on a channel and passes it on as learning information to the LSTM layers for the next learning step. LSTM layers are useful to preserve the internal state and the aggregated observations as time elapses, which helps to estimate the true state. The double $Q$-learning is used to reduce the bad states during estimation of the $Q$-value \cite{Hasselt2016}\cite{Wang2016}. Here, the user will update their DQN weights after completion of the training phase. Consistent with the requirements of a lightweight multi-user network, the implementation of the proposed algorithm is very simple. It trains the network distributively, which is executed whenever there are significant changes in the network environment.

The proposed layer architecture is shown in Fig. \ref{fig:Arch}. The architecture consists of the following layers,
\begin{itemize}
    \item Input layer: The input $x_n(t)$ is a vector of size $K+1$  where each coordinate contains the number of users which selected the given action. The input is updated during each iteration. The next iterations use a history profile (which changes according to the $Q$-values) for allocating the best channel for multi-user communication.
	\item LSTM layer: The LSTM layer \cite{SDMIA15-Hausknecht} is used to retain the internal state of the network and also helps to accumulate observations. This is needed for estimating the correct state of the network which relies on the history information. Here, the state of the network for each users' network is the load it experienced on each channel and the selected channel. In short, it learns through the experience, aggregates that experience, and passes on it over time.
	\item Value and Advantage layer: These layers help to cope with observatory problems in DQN \cite{Wang2016}. Here, we estimated the average $Q$-value of a state because every state is good or bad depending on the action taken. The average $Q$-value of an action is estimated using the $V$, the value of the state plus Adv$({a_n(t)})$, the advantage derived from the action.
	\item Output layer: It outputs a vector of size $K+1$. It consists of the estimated $Q$-value for transmission on a channel.
	\item Selection layer: The job of the selection layer is to select the best channel profile according to the $Q$-value and the load on the channel. Algorithm 1 specifies the process for the selection layer.   
	\item Distributed Double $Q$-Learning: Here, the DQN is trained using the distributed double $Q$-learning, which is used to differentiate the action from the $Q$-value \cite{Hasselt2016}\cite{Naparstek&Cohen2019}. We implemented two DQN, DQN1 for selection of the action and DQN2 for estimation of the Q-value for a given action.
\end{itemize}
\subsection{The D3RL algorithm}
\begin{figure}
\center{\includegraphics[width=0.45\textwidth]{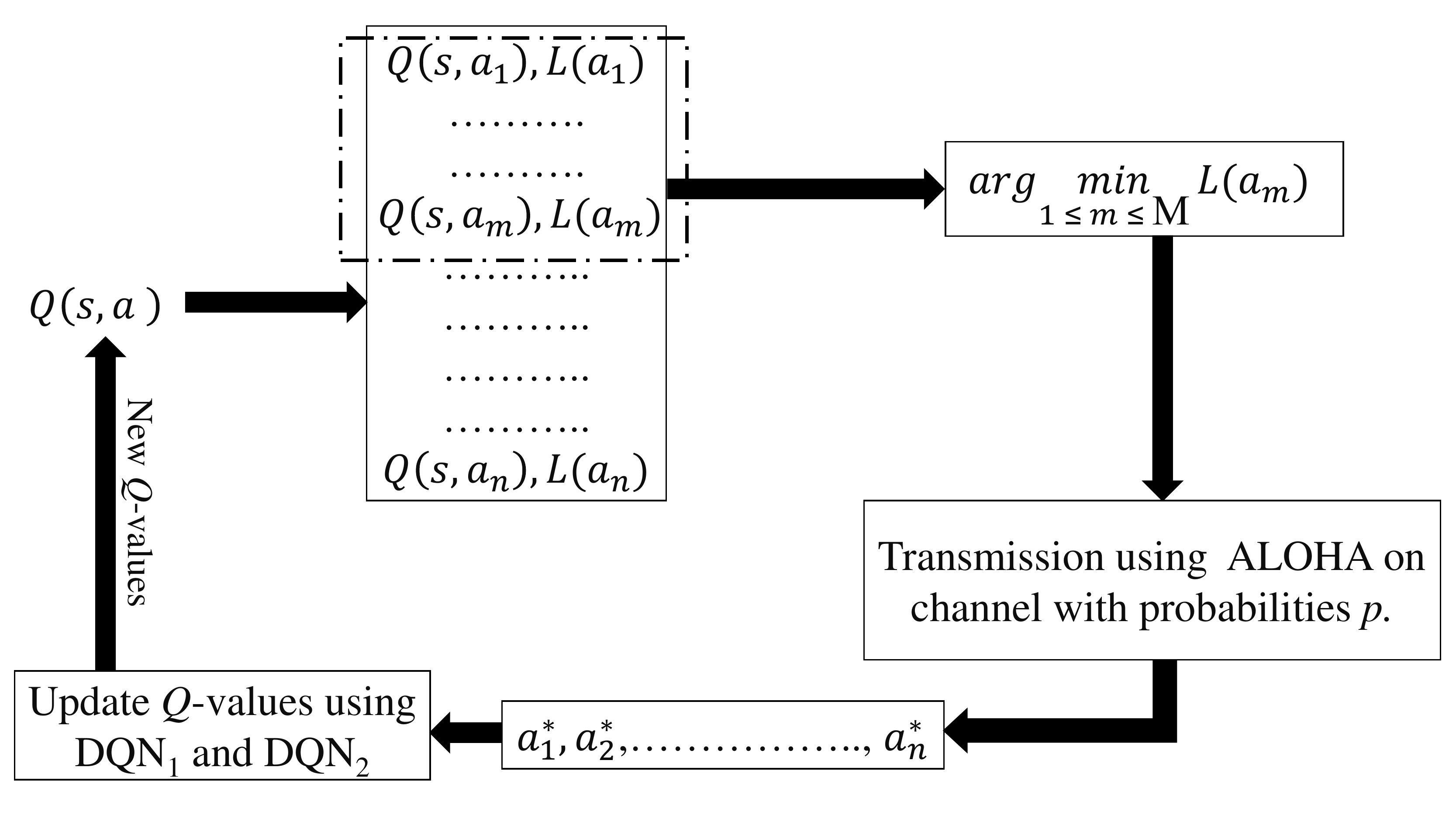}}
\caption{D3RL structure}
\label{fig:D3RL}
\end{figure}
\begin{algorithm}
\caption{Selection layer }
\KwInput{$Q(a)$, $a_n(t)$, $s_n(t)$ and $L_k$}
\KwOutput{User $n$ selects a channel profile and outputs a $Q$-value}
\For{$t \gets 1$ to $T$}
    {
        {
        \For{$k \gets 1$ to $K$}
        {
            User $n$ observes the current state $s_t$ and chooses a channel profile 
            \newline $a^*\in \arg \underset {a}\min  \{L_a: a\in a_n\},a_n \in \left\{0,\ldots,K\right\}$
                    
            Update $Q$-values for users according to the chosen channel profile.
        }
    }
}
where $a_n$=$\{a:Q(s,a)$is one of the largest $M$ values.$\}$
\end{algorithm}
Here, each user selects a channel with maximum $Q$-values out of the available $Q$-values and transmits on the channel with the least load accordingly. The flow of D3RL is shown in Fig. \ref{fig:D3RL}. The algorithm collects the $Q$-value outputs from the DQN, sorts them in descending order and considers the $M$ best channels with maximum $Q$-value, and selects the channel with the minimum load (as in Algorithm 1). In this step, user $n$ acts according to the following strategy
\begin{equation}
    \label{def:St}
    \sigma_n =\begin{cases}
    a_n = \arg  \underset {1 \leq m \leq M} \min L(a_m),    & {p}\\
    0,    & {1-p} 
    \end{cases}
\end{equation}

In the next step, the user transmits through the selected channel using ALOHA with certain transmission probabilities, generates a new action profiles for the channel and updates the $Q$-value, and pass on this new $Q$-values for the next iteration of learning.
Here, $P_{succ}$ is the success probability, $P_{NT}$ is the probability of no transmission, and $P_{coll}$ is the probability of collision. 
These probabilities are given by: 
\begin{align}
\label{def:Pnt}
     P_{NT} &= 1 - p_T \\
 \label{def:Psucc}
     P_{succ} &= p_T\left (1 - p_T\right)^{L-1} \\
 \label{def:Pcoll}
     P_{coll} &= p_T\left ( 1 - \left (1 - p_T\right)^{L-1}\right)
\end{align}
\begin{algorithm}
\caption{D3RL training}
\KwInput{Input vector $x_n(t)$}
\KwOutput{Training of DQN and output estimated $Q$-values for transmission through channels}
\For{$i \gets 1$ to $R$}
{
    \For{$e \gets 1$ to $E$}
    {
        \For{$t \gets 1$ to $T$}
        {
            \For{$n \gets 1$ to $N$}
            {
                Feed $x_n(t)$ into $DQN_1$.
                    
                Estimate the $Q$-values $Q(a)$ for all available actions $a \in \{0,1,…,K\}$.
               
                Take action $a_n(t)$ (according to Algorithm 1)  and obtain a reward $r_n{(t+1)}$.
            }
            \For{$n \gets 1$ to $N$}
            {        
                Feed $x_n{(t+1)}$ into $DQN_1$ and $DQN_2$
                    
                Estimate the $Q$-values  $\tilde{Q}_1(a)$, $\tilde{Q}_2(a)$, for all actions $a$.
           
                Construct a target vector for training by replacing the $a_n(t)$ by,
                \newline $Q(a_n(t))$ $\gets$ $r_n{(t+1)} + \tilde{Q}_2$ $(\arg \max(\tilde{Q}_1(a)))$
            }
        }
    }
    Train $DQN_1$ with $x_s$ and output $Q_s$.
    
    Every iteration set $Q_2 \gets Q_1$
    
}
\end{algorithm}
\subsection{Load Estimation}
The algorithm selects the channel with minimal load among the $M$ best channels. The load estimator uses standard ALOHA based load estimation:  
The load ($L$) on each channel is estimated using the ratio between the success probabilities and the number of transmitted packets. as the ratio of the number of successful transmissions to the total number of transmissions on the channel.
Using ~\eqref{def:Pnt}-~\eqref{def:Pcoll} this ratio is given by: 
 \begin{equation}
 \label{def:ratio}
 \frac{N_{st}^n(k)}{N_{T}^n(k)} \approx \left (1 - p_T\right)^{L-1} 
 \end{equation}
According to equation~\eqref{def:ratio}, the approximate load on the  channel is
\begin{equation}
\label{def:load}
\hat L =  1 +\left\lceil \frac {\log \left(\frac{N_{st}}{N_{T}} \right)} {\log \left( 1 - p_T\right)} \right\rceil
\end{equation}

Each time $t$ users observe a load on each of the $M$ best channels (selected according to the $Q$-values) and choose among these $M$ channels the one with the minimum load according to equation \eqref{def:St}. Here, $N_T$ is the total number of transmissions and $N_{st}$ is the  total number of successful transmissions.  
\subsection{Training Mechanism}
The DQN is trained using a D3RL training algorithm as shown below. Here, all users are trained in distributively. Training is only required when the characteristics of the network change, such as the addition of new nodes and links etc. Algorithm 2 runs for $R$ iterations where it calculates a $Q$-value for all available actions $a_n$ on the channel and learns from it during each iteration and outputs the best channel allocation for each user after $R$ iterations. Here, $E$ is the number of episodes. 

\subsection{Complexity Analysis of D3RL}
When using D3RL each user needs to find the $M$ best channels (with respect to the $Q$-values). The total time required to find the $M$ best channels has a complexity of  $ O(K\log M)$ operations per step. 

The complexity of D3RL in terms of the number of multiplications performed during the operation of the DQN is easy to compute. Assuming a DQN with $N_L$ layers, where the size of the input layer is $K$ and each of the other layers has size $n$. Therefore the real-time computational complexity (for forward and backward propagation) for every user at each time step is: 
 \begin{equation}
 \label{def:Nu}
 O\left(Kn + (N_L-1)n^2 \right).
 \end{equation}
The training phase computational complexity for each user over $R$ iterations is in order of 
 \[
 O\left(RET\left( K n+ (N_L-1)n^2 + K \log M\right)\right).
 \] 
\section{Simulation Results}
\label{SR}
In this section we present several simulated experiments comparing the proposed D3RL algorithm to DQSA \cite{Naparstek&Cohen2019}. We have implemented D3RL with $M=2,3,4,8$ channels.
The common communication and the neural network parameters used for the simulation are in Table \ref{tab:CP} and \ref{tab:NNP} respectively. 
\begin{table}
\caption{Communication Parameters}
\centering
\label{tab:CP}
\begin{tabular}{ll}
\hline
\multicolumn{1}{c}{\textit{\textbf{Parameter}}} & \multicolumn{1}{c}{\textit{\textbf{Value}}} \\
\hline
Number of users                               & 100 \\                             
Number of channels                              & 50 or 25   \\                                       
Type of channel                                 & Rayleigh Fading Channel  \\   
SNR                                             & 35dB             \\                           
Bandwidth                                       & 20MHz          \\ 


Maximal doppler shift                           & 100Hz          \\                              \hline
\end{tabular}
\end{table}

\begin{table}
\caption{Neural network parameters}
\centering
\label{tab:NNP}
\begin{tabular}{ll}
\hline
\multicolumn{1}{c}{\textit{\textbf{Parameter}}} & \multicolumn{1}{c}{\textit{\textbf{Value}}} \\ \hline
Number of LSTM layers                           & 100                                         \\
Number of advantage layers                      & 10                                          \\
Number of value layers                          & 10                                          \\
Minibatch size                                  & 16 episodes                                 \\
Time steps                                      & 10 to 100                                   \\
Discount factor($\gamma$)                       & 0.95                                        \\
Alpha factor($\alpha$)                          & 0.05                                        \\
Temperature($\beta$)                            & 1 to 20                                     \\
Number of training iterations                   & 10000                                         \\ \hline
\end{tabular}
\end{table}
\begin{figure}
\center{\includegraphics[width=8.7cm,trim={0cm 5.8cm 0cm 6cm}, clip]{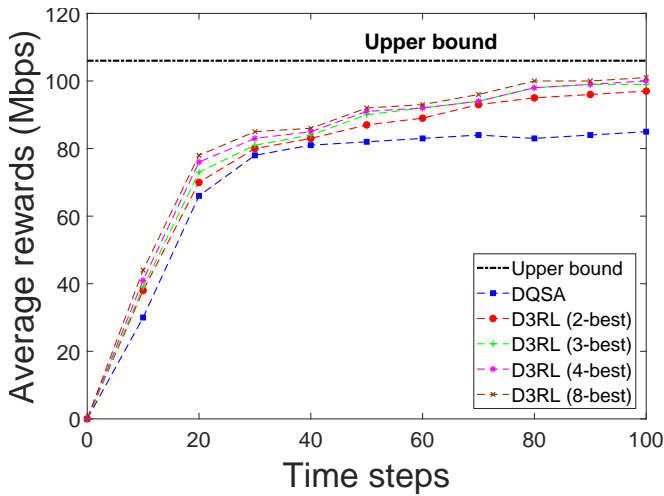}}
\caption{Comparative average reward with $i.i.d$ channel (100 users 50 channels)}
\label{fig:IID1}
\end{figure}
Fig. \ref{fig:IID1} depicts the performance of D3RL over $i.i.d$ Rayleigh fading channels. The algorithm (with $M=2$) achieves rates which are $15\%$ higher than DQSA and $18\%$ when using $M=8$. The improvement can be attributed to the deterministic learning used in the D3RL which ensures an asymptotically balanced channel allocation for each user. For comparison, we also provide an upper bound on the average rewards similar to the upper bound in  \cite{ONaparstek2012}. 

\begin{figure}

\center{\includegraphics[width=8.7cm,trim={0cm 5.8cm 0cm 6cm}, clip]{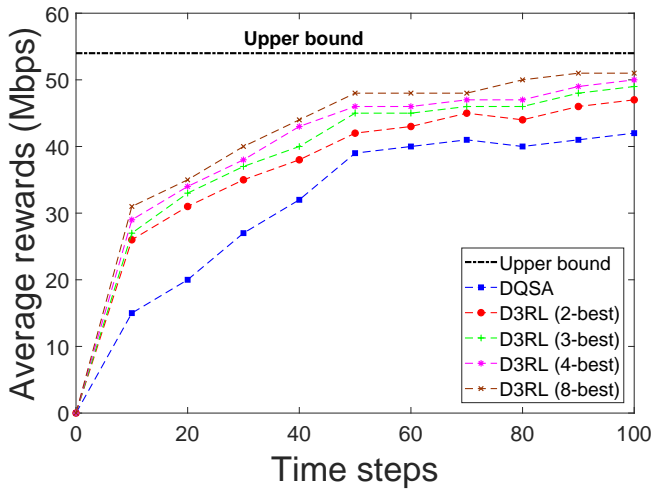}}
\caption{Comparative average reward with $i.i.d$ channel (100 users 25 channels)}
\label{fig:IID2}
\end{figure}
Fig. \ref{fig:IID2} compare the  performance of the D3RL algorithm under higher loads (100 users and 25 channels) as a function of the  time step. 
In this loaded situation, D3RL outperformed DQSA by $20\%$ (with $M=2$) and $24\%$ (with $M=8$). Fig. \ref{fig:IID2} also shows the upper bound on the average rewards on the channel. As can be seen, indeed the D3RL performance is very close to the optimal allocation as is expected by the analysis in \cite{Bistritz&Leshem2019} and by the results of \cite{Azar2000}. 

Fig. \ref{fig:IID1} and \ref{fig:IID2} also depict the performance of D3RL with $M=2,3,4,8$. As discussed above increasing $M$ improves the performance with a  minor increase in computational complexity. However, the gains are insignificant for $M>4$. This is very reasonable, since by the random graph argument of \cite{Bistritz&Leshem2019} and substituting  $M=\log K$ leads with high probability to a completely balanced NE, since in this graph there is a perfectly balanced matching with high probability.


\section{Conclusions and Future Work}
\label{Con}
We considered the problem of collaborative spectrum sharing in a multi-user ad-hoc network. We designed a distributed learning algorithm to find an asymptotically balanced channel
allocation for each user. The proposed mechanism allows each user to learn the best strategies through learning without exchanging any extra messages in the network by exploiting
the properties of the multichannel ALOHA protocol. The experimental results of the D3RL show a strong performance of the algorithm in a complex multi-user network. 

In the future, this work can be further extended to analyze system dynamics by using game theoretic techniques and developing a complete intelligent MAC mechanism exploiting DRL techniques. It can also be extended to a hardware implementation of the algorithm and testing using a hardware
test-bed.  
\bibliographystyle{IEEEbib}
\bibliography{refs}

\end{document}